\documentclass[%
prd,
endfloats*,
showpacs,
showkeys,
preprint,
nofootinbib 
] {revtex4}
\usepackage{graphicx,hyperref,amssymb,amsmath,enumerate}

\def\kd{{\rm K}^+\rightarrow \mu^+\mu^+\pi^-}
\def \lg  {\langle}
\def \rg  {\rangle}
\begin{document}
\title{Hadronic structure aspects of $K^+\to \pi^-+ l^{+}_1 + l^{+}_2$ decays}
\date{\today}

\author{Mikhail A. Ivanov}
\email{ivanovm@thsun1.jinr.ru}
\affiliation{Joint Institute for Nuclear Research, Dubna,
Russia}

\author{Sergey G. Kovalenko}
\email{sergey.kovalenko@usm.cl}
\affiliation{Departamento de F{\' \i}sica, Universidad T{\' e}cnica Federico
Santa Mar{\' \i}a, Casilla 110-V, Valpara{\' \i}so, Chile}

\begin{abstract}
As is known from previous studies the lepton number violating decays
$K^+\rightarrow \pi^- + l^{+}_1 + l^{+}_2$ have
good prospects to probe new physics beyond the Standard Model
and provide valuable information on neutrino masses and mixing.

We analyze these processes with an emphasis on their hadronic structure aspects
applying relativistic constituent quark model.
We conclude that the previously ignored contribution associated with the t-channel
Majorana neutrino exchange is comparable with the s-channel one in a wide range of neutrino
masses. 
We also estimated model independent absolute upper bounds on 
the neutrino contributions to these decays. 
 \end{abstract}

\pacs{11.30.Fs, 13.20.Eb, 12.39.-x, 14.60.Pq, 14.60.St}

\keywords{lepton number, meson, neutrino}

\maketitle

\section{\label{sec:introduction}Introduction}


Discovery of small but finite neutrino masses and large neutrino flavor mixing
has clearly shown the limitations of the Standard model (SM) of electroweak
interactions and pointed to the physics beyond its framework.
The smallness of neutrino masses is commonly considered as a strong
indication in favor of the celebrated seesaw picture \cite{seesaw} with its characteristic
attributes: high-energy scale of lepton number violation (LNV) associated with
new physics as well as very light and very heavy Majorana neutrino mass eigenstates 
(for a recent review see, for instance, Refs. \cite{seesaw-recent}).
This supports the long standing belief that, contrary to the SM,
lepton number is not conserving and neutrinos are massive Majorana particles.
If this is true, the LNV processes, forbidden in the SM, are allowed at small rates
and some of them can be observed experimentally.
Therefore, theoretical studies and experimental searches for LNV processes are attracting growing
interest as the way to probe new physics beyond the SM and to study the properties of neutrino.

Various LNV processes have been discussed in the literature in this respect
(for review see~\cite{rod02,Dib:2000ce}). In principle, they can probe Majorana
neutrino contribution and provide information on the so called effective Majorana mass
matrices $\langle m_\nu \rangle_{\alpha\beta}$ and $\langle M_{N}^{-1} \rangle_{\alpha\beta}$ of
light and heavy Majorana neutrinos.
These quantities under certain assumptions are related to the
entries of the Majorana neutrino mass matrix $M_{\alpha\beta}^{(\nu)}$.
Currently the most sensitive experiments intended to probe LNV physics beyond the SM, 
in particular, Majorana neutrino contribution are those searching
for nuclear neutrinoless double beta ($0\nu\beta\beta$) decay~\cite{kla01a,aal02,arn04}. Due to the lepton flavor
structure of this process its experimental searches are sensitive to a specific
flavor set of the LNV parameters. For the Majorana neutrino contribution to this
process they are $\langle m_\nu \rangle_{ee}$ and $\langle M_{N}^{-1} \rangle_{ee}$ entries
of the effective Majorana neutrino mass matrices.  In order to probe the LNV
parameters with another lepton flavor composition one needs to study other
LNV processes.

In the present paper we study LNV  $K^+\rightarrow \pi^-  l^{+}_1  l^{+}_2$ decays. Currently the best
experimental upper bounds on the branching ratios of these processes are \cite{PDG}
\begin{eqnarray}\label{kdec-lim1}
{\cal R}_{\mu\mu} &=& \frac{\Gamma(\kd)}{\Gamma({\rm K}^+\rightarrow all)}
\leq 3.0\times 10^{-9}, \ \ \ 
{\cal R}_{e e} = \frac{\Gamma(K^+\rightarrow \pi^- e^{+} e^{+})}{\Gamma({\rm K}^+\rightarrow all)}
\leq 6.4\times 10^{-10},\\ \nonumber
{\cal R}_{\mu e} &=& \frac{\Gamma(K^+\rightarrow \pi^- \mu^{+} e^{+})}{\Gamma({\rm K}^+\rightarrow all)}
\leq 5.0\times 10^{-10}.
\end{eqnarray}
These processes may receive various contributions from the LNV physics beyond the SM 
(see, for instance, Ref. \cite{dorokhov:2002}), including the
Majorana neutrino exchange. We concentrate on the latter case.

Assume the neutrino mass spectrum consists of light $\nu_k$ and heavy $N_k$
neutrinos with the masses much smaller $m_{\nu(k)}\ll m_K$ and much larger
$M_{N(k)}\gg m_k$ than the K-meson mass $m_K=494$MeV respectively. Then the light
and heavy neutrino contributions to the amplitude of $K^+\rightarrow \pi^- l^{+}_1 l^{+}_2$ 
decay are proportional to the effective masses
$\langle m_\nu \rangle_{l_1l_2}$ and $\langle M_{N}^{-1} \rangle_{l_1l_2}$ with $l_i=e, \mu$ .
The estimates of these quantities (see, for instance, 
~\cite{Dib:2000wm,rod02})
from the neutrino observations lead to the so small branching ratios of these decays
in comparison with the current experimental sensitivities (\ref{kdec-lim1})
that their experimental observation 
looks unrealistic even in a distant future. The exception occurs if
there exists Majorana neutrino $\nu_h$ with the mass in the ``resonant" region. For the $\kd$ decay this 
is the region of 245 MeV$ \leq m_{\nu_h} \leq  388$ MeV. In this case the $\nu_h$ contribution 
is resonantly enhanced and may result in an observable effect ~\cite{Dib:2000wm}.

In the SM extensions with Majorana neutrinos there are two
lowest order diagrams, shown in Fig.1, which contribute 
to the $K^+\rightarrow \pi^- l^{+}_1 l^{+}_2$ decays.
These diagrams were first considered for $\kd$ decay long ago in 
Refs.~\cite{Ng:1978ij,Abad:1984gh} and more recently in Ref. \cite{Kdec3}.

The contribution from the factorizable s-channel diagram in
Fig.~1(a), dominant for the neutrinos with the masses in the
``resonant" region, can be calculated without referring to any
hadronic structure model. On the contrary the t-channel diagram in
Fig.~1(b) requires a detailed hadronic structure calculation.
Studying neutrino contributions to $K^+\rightarrow \pi^- l^{+}_1 l^{+}_2$ decays outside the
``resonant" region one should take into account both diagrams.
This implies the analysis based on a certain model of hadronic structure.


\begin{figure}
  \begin{center}
    \includegraphics[width=1\textwidth,height=0.25\textheight]{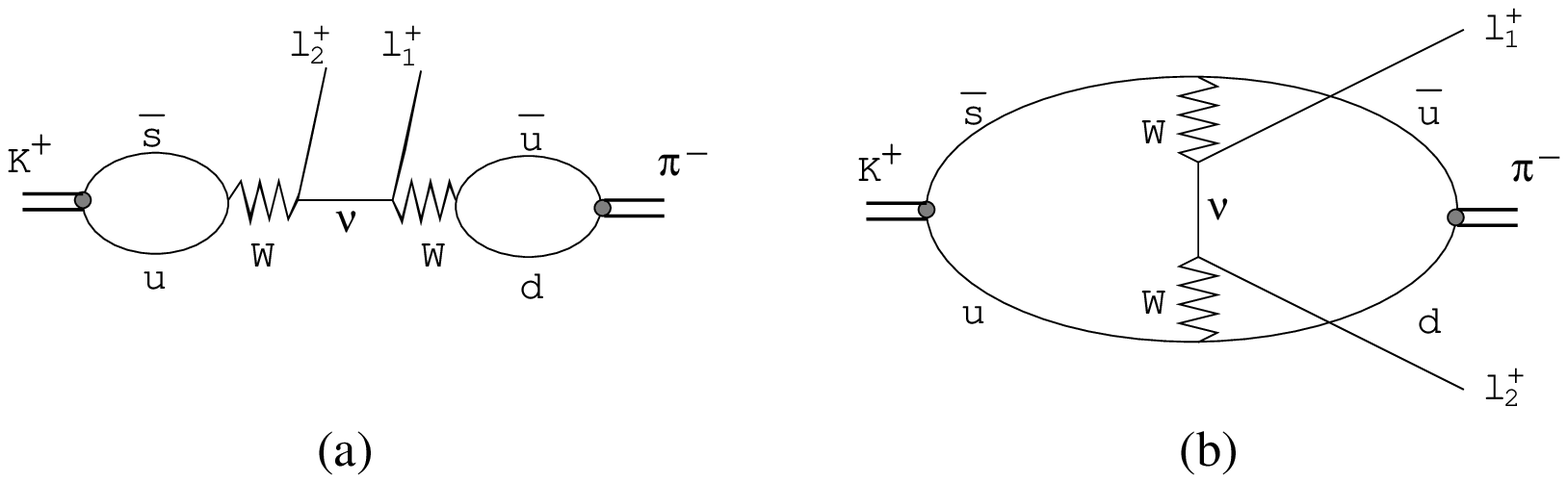}
  \end{center}
  \caption{The lowest order diagrams contributing to $K^+\rightarrow \pi^- l^{+}_1 l^{+}_2$ decays.  
\label{fig.1}}
\end{figure}

In what follows we focus on the hadronic structure aspects of $K^+\rightarrow \pi^- l^{+}_1 l^{+}_2$ decay.
One of the main motivation of our study is the controversial situation existing
in the literature on this subject.
In Ref. \cite{Abad:1984gh} the contribution of the t-channel diagram in
Fig.~1(b) has been evaluated in the Bethe-Salpeter approach and
argued to be negligible compared to
the s-channel diagram in Fig.~1(a) for any value of neutrino
mass. To our mind this result is not supported by any physical
reason and looks as an artefact of this approach. In this
situation it is worthwhile to carry out an independent analysis of the
t-channel contribution within an alternative approach to hadronic
structure calculations.

Our analysis is based on the relativistic constituent quark 
model~\cite{model}
which was successful
in description of various meson decay processes.
As will be demonstrated, we disagree with the above mentioned conclusion of 
Ref.~\cite{Abad:1984gh} and
predict that the t-channel neutrino contribution to $K^+\rightarrow \pi^- l^{+}_1 l^{+}_2$ decays is comparable
with the s-channel one for all the values of neutrino masses outside the ``resonant" region.

The following comment is in order. In view of the fact mentioned before that neutrinos with
the masses outside the ``resonant" region
give an experimentally undetectable contribution to $K^+\rightarrow \pi^- l^{+}_1 l^{+}_2$ decays 
the significance of
our results mainly consists in establishing a reliable framework for hadronic
structure calculations in the analysis of these and similar exotic decays rather than
for extraction of neutrino parameters. On the other hand one may note that our results,
obtained for the simplest neutrino exchange mechanism of 
$K^+\rightarrow \pi^- l^{+}_1 l^{+}_2$ decays, can be straightforwardly extended
to some other mechanisms offered by the physics beyond the SM, which could lead to {\it a priori} 
much larger rates and provide valuable information on the LNV physics.


\section{\label{sec:1}Majorana neutrino contribution to $K^+\to \pi^- l^{+}_1  l^{+}_2$ decays}

We consider the SM extension with massive Majorana neutrinos.
In this case the weak interaction effective Lagrangian has
the standard form. For the studied processes the relevant terms are
\begin{equation}\label{Lagr}
{\cal L}_{\rm int}^{\rm weak} =\frac{G_F}{\sqrt{2}}
\left[V_{ud}\,\bar d O^\alpha u + V_{us}\,\bar s O^\alpha u \right]\cdot
\bar\nu_k U^*_{n k} O_\alpha l_n + {\rm h.c.}
\end{equation}
with $O^\alpha=\gamma^\alpha(1-\gamma^5)$.
The unitary neutrino mixing matrix $U_{ij}$ relates $\nu'_i = U_{ik}\nu_k$
weak $\nu'_i$ and Majorana neutrino mass  eigenstates $\nu_k$ with 
the masses $m_{\nu_k}$. The fields $l_n$ denote charged leptons  $e$, $\mu$ and $\tau$. 

The lowest order diagrams describing Majorana neutrino
contribution to \mbox{$K^+(p)\to \pi^-(p')+l_1^+(q_1)+l_2^+(q_2)$}
decays are shown in Fig.~1. The corresponding matrix elements
we write down in the form
\begin{eqnarray}
\label{kpi} \nonumber
M(l_1^+ l_2^+)&=& G_F^2\, V_{us} V_{ud}\,
\sum_k 
U_{l_1k}U_{l_2k}m_{\nu_k}
\left[H^{\mu_1\mu_2}(q_1,q_2;m_{\nu_k})\cdot L_{\mu_1\mu_2}(q_1,q_2) -(q_1\leftrightarrow q_2)\right]=\\
&=& M(l_1^+ l_2^+)_s + M(l_1^+ l_2^+)_t.
\end{eqnarray}
Here the terms $M(l_1^+ l_2^+)_s$ and $M(l_1^+ l_2^+)_t$ denote the contributions of the s- and t-channel diagrams
in Fig. 1(a) and Fig.1(b) respectively.
The lepton and hadron tensors are defined as 
\begin{eqnarray}
\label{lept-tens}
L_{\mu_1\mu_2}(q_1,q_2)&=&
l_1^{\lambda_1,+}(q_1) C\gamma_{\mu_1}\gamma_{\mu_2}
(1-\gamma_5)l_2^{\lambda_2,+}(q_2),
\\
H^{\mu_1\mu_2}(q_1,q_2;m_\nu) &=& 
    H^{\mu_1\mu_2}_s(q_1,q_2;m_\nu)
 +  H^{\mu_1\mu_2}_t(q_1,q_2;m_\nu).
\label{had-tens}
\end{eqnarray}
Here $\lambda_1$, $\lambda_2$ are the polarizations of the charged leptons and $C$ is the charge conjugation matrix.
The hadron matrix elements $ H^{\mu_1\mu_2}_s$ and $ H^{\mu_1\mu_2}_t$
correspond to the contributions of the s- and t-channel diagrams in Fig. 1(a) and Fig.1(b) respectively.

The contribution of the factorizable s-channel diagram in
Fig.~1(a) can be calculated in a straightforward way without
referring to any hadronic structure model with the following result
\begin{eqnarray}\label{s-ch-def}
 H^{\mu_1\mu_2}_s(q_1,q_2;m_\nu) = 
p^{\mu_1}\, p^{\prime \mu_2}
\frac{f_\pi f_K}{m^2_{\nu}-(p-q_1)^2},
\end{eqnarray}
where the pion $f_{\pi}$ and K-meson $f_{K}$ leptonic decay constants 
completely parameterize the hadronic structure of this
contribution. Their experimental values are $f_{\pi}=131$MeV 
and $f_{K}=161$MeV. The t-channel diagram in Fig.1(b) is much more 
involved and requires calculations on the basis of certain model of 
hadronic structure. In the following sections we apply for this purpose 
the relativistic constituent quark model \cite{model}.

Let us note that for the case of neutrino mass spectrum consisting of very 
light, $m_{\nu}\ll m_K$, 
and very heavy, $m_{\nu}\equiv M_N\gg m_K$, neutrinos ($m_K=493.677$ MeV is the K-meson mass) 
both s- and t-channel matrix elements
in Eq.~(\ref{kpi}) are reduced to the form
\begin{eqnarray}\label{decomp}
M(l_1^+ l_2^+)_{s,t} = 
\frac{\langle m_{\nu}\rangle_{l_1l_2}}{m_K}{\cal A}_{s,t}^{(\nu)} +
\langle M_N^{-1}\rangle_{l_1l_2} m_K{\cal A}_{s,t}^{(N)}
\end{eqnarray}
with the contributions proportional to the effective Majorana masses
of the light $\nu$ and heavy $N$ neutrinos defined as
\begin{eqnarray}\label{average}
\lg m_{\nu}\rg_{l_1l_2} = \sum_{k=light} U_{l_1 k}U_{l_2 k} m_{\nu_k},
\ \ \
\lg M_N^{-1}\rg_{l_1l_2}  =  \sum_{k=heavy} U_{l_1 k}U_{l_2 k} M_{N_k}^{-1}.
\end{eqnarray}
In this limiting case the coefficients ${\cal A}_{s,t}^{(\nu)}$ and ${\cal A}_{s,t}^{(N)}$ are independent 
of neutrino masses and mixing. As follows from Eq. (\ref{s-ch-def}) the coefficients 
${\cal A}_{s}^{(\nu)}, {\cal A}_{s}^{(N)}$ do not depend on hadronic structure model and their values can be easily
calculated (see, for instance, Refs. \cite{Dib:2000wm,rod02}). 
We will show in Sec.~\ref{sec:3} that ${\cal A}_{t}^{(N)}$ is also hadronic model independent. 
Thus, the only coefficient in Eq. (\ref{decomp}) which requires hadronic model based calculation is ${\cal A}_{t}^{(\nu)}$.

\section{\label{sec:2}Formalism of hadronic structure calculations}

Here we present the details of the relativistic constituent quark
model \cite{model} which we apply to the calculation of 
\mbox{$K^+\to l_1^+ l_2^+ \pi^-$} decay rates.
The model is based on the effective interaction Lagrangian describing
the couplings between hadrons and their constituent quarks.
The coupling of a meson $H(q_1\bar q_2)$ to its constituent
quarks $q_1$ and $\bar q_2 $ is given by the Lagrangian
\begin{equation}\label{Lagr_str}
{\cal L}_{\rm int}^{\rm Str}(x) = g_H H(x)\int\!\! dx_1 \!\!\int\!\!
dx_2 F_H (x,x_1,x_2)\bar q(x_1)\Gamma_H\lambda_H q(x_2) \, + {\rm h.c.}
\end{equation}
Here, $\lambda_H$ and $\Gamma_H$ are the flavor $SU(3)$ Gell-Mann matrix and certain combination of 
Dirac $\gamma$-matrices corresponding to the flavor and spin quantum numbers of the meson field
$H(x)$. The function $F_H$ is related to the scalar part of the
Bethe-Salpeter amplitude and characterizes the finite size of the
meson. 
The translational invariance requires the vertex function $F_H$ to fulfil the identity 
\begin{eqnarray}\label{iden}
F_H(x+a,x_1+a,x_2+a)=F_H(x,x_1,x_2)\ \ \  \mbox{for any 4-vector}\ \  a_{\mu}.
\end{eqnarray}
We use for this function the following form 
\begin{equation}\label{vertex}
F_H(x,x_1,x_2)=\delta(x - c_{1}x_1 - c_{2}x_2) \Phi_H((x_1-x_2)^2)
\end{equation}
where $\Phi_H$ is the correlation function of the two constituent quarks
with the masses $m_1$, $m_2$. Here we introduced the notation: $c_{i}=m_i/(m_1+m_2)$. 
The form of the vertex function in Eq. (\ref{vertex}) implies factorization of 
its dependence on the center-of-mass coordinate 
\mbox{$x=(m_1/(m_1+m_2))\,x_1+(m_2/(m_1+m_2))\,x_2$} of the constituent quarks. 

The interaction Lagrangian for the particular case of charged kaon and pion takes the form
\begin{eqnarray}
{\cal L}_{\rm int}^{\pi,K}(x) &=&
ig_K K^+(x)\int\!\! dx_1 \!\!\int\!\,dx_2
F_K (x,x_1,x_2)\bar u(x_1)\gamma^5 s(x_2)
\label{Lagr_pik}\\
&+&ig_\pi \pi^+(x)\int\!\! dx_1 \!\!\int\!\,dx_2
F_\pi (x,x_1,x_2)\bar u(x_1)\gamma^5 d(x_2) \, + {\rm h.c.}
\nonumber
\end{eqnarray}

The coupling constants $g_H$ in Eqs.~(\ref{Lagr_str}), (\ref{Lagr_pik}) are determined by the
so-called {\it compositeness condition} 
which requires the renormalization constant of
an elementary meson field $H(x)$ to vanish
\begin{equation}\label{z=0}
Z_H \, = \, 1 - \, \frac{3g^2_H}{4\pi^2} \,
\tilde\Pi^\prime_H(M^2_H) \, = \, 0,
\end{equation}
where $\tilde\Pi^\prime_H$ is the derivative of the meson 
self-energy function.
In the case of pseudoscalar mesons we have
\begin{eqnarray}\label{Mass-operator}
\tilde\Pi^\prime_P(p^2) \, = \frac{1}{2p^2}\,
p^\alpha\frac{d}{dp^\alpha}\,
\int\!\! \frac{d^4k}{4\pi^2i} \tilde\Phi^2_P(-k^2)
{\rm tr} \biggl[\gamma^5 S_1(\not\! k+c_{1} \not\!p)
                \gamma^5 S_2(\not\! k-c_{2} \not\!p) \biggr] \, ,
\end{eqnarray}
where $\tilde\Phi_P(-k^2)$ is the Fourier-transform of the
correlation function  $\Phi_P((x_1-x_2)^2)$ and
$S_i(\not\! k)$ is the quark propagator.
We use the free fermion propagators for the valence quarks
\begin{equation}
S_i(\not\! k)=\frac{1}{m_i-\not\! k}
\end{equation}
with an effective constituent quark mass $m_i$.
In order to avoid the appearance of the imaginary parts in the
physical amplitudes we require
\begin{equation}
\label{conf}
M_P < m_{1} + m_{2}
\end{equation}
for the meson mass $M_P$.

Finally we specify the correlation function
$\Phi_H$ in Eqs.~(\ref{vertex}), (\ref{Mass-operator}) characterizing finite
size of hadrons. Any choice for its Fourier-transform $\tilde\Phi_H$ is appropriate
as long as it falls off sufficiently fast in the ultraviolet region of
the Euclidean space to render Feynman diagrams ultraviolet finite.
We adopt the Gaussian form for this function
\begin{eqnarray}\label{vertex-f}
\tilde\Phi_H(k^2_E) \doteq \exp(- k^2_E/\Lambda^2_H),
\end{eqnarray}
where $k_E$ is Euclidean momentum. The hadronic size parameters 
$\Lambda_H$ and the constituent quark masses
$m_{u,d,s}$ are determined by fitting to the experimental data for the leptonic
decay constants $f_H$ of mesons $H$.
The model expressions for the leptonic decay constants $f_P$ of 
pseudoscalar mesons are derived
from the Lagrangian (\ref{Lagr_pik}) and take the form
\begin{equation}
\label{fP}
{\cal F}_P(p^2) \,p^\mu \, = \, \frac{3g_P}{4\pi^2}
\,\int\!\! \frac{d^4k}{4\pi^2i}
\tilde\Phi_P(-k^2) {\rm tr} \biggl[O^\mu S_1(\not\! k+c_{1} \not\!p)
\gamma^5 S_2(\not\! k-c_{2} \not\!p) \biggr] \,
\end{equation}
with the definition $f_P \doteq {\cal F}_P(M_P^2)$.
The best fit to the experimental values of the decay constants $f_\pi=131$ MeV and
$f_K=161$ MeV is obtained with
\begin{eqnarray}
m_{u(d)} &=& 0.235\ \mbox{GeV} \hspace{1cm} m_s=0.333\ \mbox{GeV}
\label{fit}\\
\Lambda_\pi &=& 1.0\ \mbox{GeV} \hspace{1.5cm} \Lambda_K=1.6\ \mbox{GeV}
\nonumber
\end{eqnarray}

This completes the definition of the model which we apply to the analysis of
$K^+\to l_1^+  l_2^+ \pi^-$ decays.

\section{\label{sec:3} $K^+\to l_1^+  l_2^+ \pi^-$ hadronic matrix elements and decay rates}

Now let us turn to the calculation of the hadronic matrix elements of
$K^+\to l_1^+  l_2^+ \pi^-$ decays within the above presented
approach. The Lagrangian describing these processes consists of the three terms
\begin{eqnarray}\label{full-Lag}
{\cal L}_{K-dec} = {\cal L}_{\rm int}^{\rm weak} + {\cal L}_{\rm int}^{\pi}+
{\cal L}_{\rm int}^{K},
\end{eqnarray}
where the first term is the weak interaction Lagrangian (\ref{Lagr}) while
the second and the third terms determine $\pi$ and $K$ meson interactions with
quarks defined in Eq. (\ref{Lagr_pik}). In the lowest order this Lagrangian
generates the contributions corresponding to the diagrams in Fig.~1.
In what follows we concentrate on the contribution of the t-channel
diagram in Fig.~1(b). The expression for the corresponding 
hadronic matrix element introduced in Eqs.~(\ref{kpi}), (\ref{had-tens})
takes the form
\begin{eqnarray}
\label{t-ch-model}
 H^{\mu_1\mu_2}_t (q_1,q_2;m_{\nu})&=& -\,3 g_\pi g_K
\int\frac{d^4k_1}{(2\pi)^4i}\int\frac{d^4k_2}{(2\pi)^4i}
\tilde\Phi_K(-k_1^2)\tilde\Phi_\pi(-k_2^2)
\\
&\times&
{\rm tr}
\left[\gamma^5\, S_s(k_1-c_2 p)\,O^{\mu_1}\, S_u(k_2-p'/2)\,\gamma^5\,
S_d(k_2+p'/2)\,O^{\mu_2}\, S_u(k_1+c_1 p)\right]
\nonumber\\
&\times&
\frac{1}{m^2_{\nu}-(k_1-k_2+q_{12})^2}
\nonumber
\end{eqnarray}
where $q_{12}=c_1q_1-c_2q_2+(1/2-c_2)p'$ with $c_1=m_u/(m_u+m_s)$ and $c_2=m_s/(m_u+m_s)$.
The sign minus comes from one fermion loop.

We note that the characteristic energy scale of $K^+\to l_1^+l_2^+\pi^-$ is set by $m_K$. 
Therefore for the neutrino masses $m_\nu\gg m_K$ the neutrino
propagators in the matrix elements of these processes can be substituted by the constant
$$
\frac{1}{m^2_\nu-k^2}\to \frac{1}{m^2_\nu}
$$
Thus the direct dependence on the final lepton momenta $q_1$ and $q_2$ drops out from
the invariant matrix elements in Eqs.~(\ref{s-ch-def}) and (\ref{t-ch-model}). 
Using the Fierz identity 
\begin{equation}\label{Fierz}
 {\rm tr}\left(T_1 O^\mu T_2 O_\mu\right)=
-{\rm tr}\left(T_1 O^\mu\right) {\rm tr}\left(T_2 O_\mu\right)
\end{equation}
in Eq.~(\ref{t-ch-model}) and recalling the definition of the
weak decay constants $f_\pi$ and $f_K$ in Eq.~(\ref{fP}), one finds that
\begin{equation}
\label{infinity1}
H^{\mu_1\mu_2}_t(q_1,q_2;m_\nu)=
\frac{1}{3} H^{\mu_1\mu_2}_s(q_1,q_2;m_\nu)\ \ \ \ \mbox{for}\ \ m_\nu\gg m_K.
\end{equation}
Thus, in this limit the t-channel contribution can be evaluated in a hadronic model independent way as well as
the s-channel contribution. 

In the case of arbitrary finite neutrino masses,
after straightforward  but  quite tiresome calculations, 
explained in Appendix~\ref{sec:appA},
we end up with the following expression for the t-channel hadronic matrix element
\begin{eqnarray}\nonumber
H^{\mu_1\mu_2}_t (q_1,q_2;m_\nu) &=&
H_t^{g}(s_1,s_2;m_{\nu})\,g^{\mu_1\mu_2}
+ H_t^{p'p}(s_1,s_2;m_{\nu})\, \left(p^{\mu_1}\,p^{\prime\,\mu_2}+p^{\prime\,\mu_1}\,p^{\mu_2} +
i\varepsilon^{\rho_1\rho_2\mu_1\mu_2}\,p_{\rho_1}\,p^{\prime}_{\rho_2} \right)\\
\label{app-f}
&+& H_t^{p'q}(s_1,s_2;m_{\nu})\,
\left(p^{\prime\,\mu_1}\,q_{12}^{\mu_2}+q_{12}^{\mu_1}\,p^{\prime\,\mu_2} -
i\varepsilon^{\rho_1\rho_2\mu_1\mu_2}\,p'_{\rho_1}\,q_{12\,\rho_2}\right).
\end{eqnarray}
An approximate analytic representation for the structure functions
$H_t^{k}(s_1,s_2;m_{\nu})$ is given in (\ref{approx}).
We define the kinematical variables as
\begin{eqnarray*}
s_1 &=& (q_1+q_2)^2=(p-p')^2,
\\
s_2 &=& (q_2+p')^2=(p-q_1)^2,
\\
s_3 &=& (p'+q_1)^2=(p-q_2)^2,
\end{eqnarray*}
where $p^2=m^2_K$, $p^{\prime\,2}=m^2_\pi$, $q^2_i=m^2_{l_i}$ and
$s_1+s_2+s_3=m^2_K+m^2_\pi+m^2_{l_1}+m^2_{l_2}$.

With these definitions the $K^+\to l_1^+ l_2^+ \pi^-$ decay rate can be written in 
the form
\begin{eqnarray}
\label{width1}
\Gamma(K^+\to l_1^+ l_2^+ \pi^-) &=&
\left(1 - \frac{\delta_{l_1l_2}}{2}\right)\frac{G^4_F V^2_{us} V^2_{ud} }{256\pi^3m_K^3}
\sum_{k,n} 
\alpha^{l_1l_2}_{kn}
\int\limits_{(m_{l_1}+m_\pi)^2 }^{(m_K-m_{l_2})^2}\! ds_3
\int\limits_{s_2^-}^{s_2^+}\! ds_2 \,{\cal F}(s_2,s_3)_{kn},
\end{eqnarray}
where $\alpha^{l_1l_2}_{kn} = (U_{l_1 k}U_{l_2 k} m_{\nu_k})(U^*_{l_1 n}U^*_{l_2 n} m_{\nu_n})$ and 
\begin{eqnarray}\label{def-1}
s_2^\pm &=& m^2_{l_1}+m^2_{K}-\frac{1}{2s_3}
\left[
(s_3+m^2_K-m^2_{l_2})(s_3+m^2_{l_1}-m^2_\pi) \mp\right.\\ \nonumber
&&\ \ \ \ \ \ \ \ \ \ \ \ \ \ \ \ \ \ \ \ \ \ \ \ \left.\mp \lambda^{1/2}(s_3,m^2_K,m^2_{l_2})\lambda^{1/2}(s_3,m^2_{l_1},m^2_\pi)
\right],
\end{eqnarray}
The integrand in Eq. (\ref{width1}) is
\begin{eqnarray}\nonumber
{\cal F}( s_2,s_3)_{kn} &=& 
2\left(
H^{\mu_1\mu_2}(q_1,q_2;m_{\nu_k})+H^{\mu_2\mu_1}(q_2,q_1;m_{\nu_k})
\right)\,
\left(
 H^{\dagger\,\nu_1\nu_2}(q_1,q_2;m_{\nu_n})
+H^{\dagger\,\nu_2\nu_1}(q_2,q_1;m_{\nu_n})
\right)\,\\
&&\times q_1^{\rho_1} q_2^{\rho_2} {\rm   tr}\gamma_{\mu_1}\gamma_{\mu_2}\gamma_{\rho_2}\gamma_{\nu_2}\gamma_{\nu_1}\gamma_{\rho_1}(1-\gamma^5).
%
\end{eqnarray}
An explicit form of the function ${\cal F}(s_2,s_3)_{kn}$, which we do not show here for its complexity, 
is derived by the substitution of the expression for $H^{\mu_1\mu_2}$ from Eqs. (\ref{had-tens}), (\ref{s-ch-def}) 
and (\ref{app-f}). Then we carry out the twofold integration in Eq. (\ref{width1}) numerically. The results of 
these calculations we discuss in the next section.

\section{\label{section:numerical} Numerical results}

One of the main purposes of our study is to examine the relative contribution of the
t-channel diagram,  Fig~1(b), to $K^+\to l_1^+ l_2^+ \pi^-$ decays.
We are doing this in terms of the decay rates of these processes 
comparing their values obtained in the case when
both s- and t-channel diagrams are taken into account, $\Gamma_{s+t}$, with the case 
when the t-channel diagram is switched off, $\Gamma_{s}$.
For the sake of simplicity we analyze the contribution of only one neutrino mass eigenstate $\nu_h$ 
with an arbitrary mass $m_{\nu_h}$. 
Varying  $m_{\nu_h}$ in a wide range of its values we assume $\nu_h$
to be an additional mass eigenstate to the three ordinary very light neutrinos. This additional neutrino state 
may appear in models with sterile neutrino species (see, for instance, Refs. \cite{Dib:2000wm,GKS:2001}) 
and may a priori have an arbitrary mass.

We present our results for the particular case of $\kd$ decay in Table~1 for the total decay rate 
$\Gamma_{s+t}/|U_{\mu h}|^4$ and for the ratio $\Gamma_{s+t}/\Gamma_{s}$ as
functions of neutrino mass $m_{\nu_h}$. For other decays $K^+\to l_1^+ l_2^+ \pi^-$ the results are similar.

The following comments are in order. The s-channel diagram has the two
singular neutrino propagators \mbox{$1/(m^2_{\nu_h}-s_{2,3})$}.
Therefore, for the neutrino mass $m_{\nu_h}$  in the ``resonant" intervals
\begin{eqnarray}
\label{domain}
(m_{\mu}+m_{\pi}) \approx 245 \mbox{ MeV} &\leq& 
m_{\nu_h} \leq (m_K-m_{\mu}) \approx 388 \mbox{ MeV}, \ \ \ \mbox{for} \ \ \ K^+\to \mu^+ \mu^+ \pi^-,\\
\nonumber
(m_{e}+m_{\pi}) \approx 140 \mbox{ MeV} &\leq& 
m_{\nu_h} \leq (m_K-m_{e}) \approx 493 \mbox{ MeV}, \ \ \ \mbox{for} \ \ \ K^+\to e^+ e^+ \pi^-, e^+ \mu^+ \pi^- 
\end{eqnarray}
one must take into account the total decay width $\Gamma_{\nu_h}$ of 
$\nu_h$-neutrino substituting \mbox{$m_{\nu_h}\rightarrow m_{\nu_h} - (i/2)\Gamma_{\nu_h}$.}
The total decay width $\Gamma_{\nu_h}$ receives all the possible
contributions from the leptonic and semi-leptonic
charged and neutral current decay modes allowed by the energy-momentum 
conservation for the Majorana neutrino $\nu_h$ with the mass in the
resonant intervals (\ref{domain}). For the resonant interval of the $\kd$ decay this quantity has been calculated in
Refs.~\cite{Dib:2000wm,GKS:2001} as a function of $m_{\nu_h}$. 
In our analysis we use its average value over the resonant
intervals (\ref{domain}) which is  $\Gamma_{\nu_h}\approx 10^{-14}$GeV.

Due to the smallness of $\Gamma_{\nu_h}$ the
s-channel diagram in Fig.~1(a) blows up in the resonant intervals and absolutely
dominates over the t-channel one. This effect is
clearly seen in Table~1. 

Let us note that the values of the decay rates 
in the resonant intervals (\ref{domain}), dominated by $\Gamma_{s}$, 
give model independent theoretical upper limits on the neutrino 
contributions to the studied processes
\begin{eqnarray} \label{upper-bound}
\Gamma(K^+\to l_1^+ l_2^+ \pi^-)\leq \Gamma_s^{res}(K^+\to l_1^+ l_2^+ \pi^-). 
\end{eqnarray}
The t-channel contributions introduce negligible corrections to these limits. 
For the corresponding branching ratios we obtain the following order of magnitude upper limits
\begin{eqnarray} \label{upper-bound1}
{\cal R}_{\mu\mu}, {\cal R}_{ee}, {\cal R}_{\mu e}\leq 10^{-1}. 
\end{eqnarray}
The derivation of more accurate limits requires a comprehensive evaluation of the total decay width $\Gamma_{\nu_h}$
of the neutrino mass eigenstate $\nu_h$ as a function of $m_{\nu_h}$ in the resonant intervals (\ref{domain}), which
is beyond the scope of the present study.
Apparently the limits in Eq. (\ref{upper-bound1}) are much larger than the experimental limits 
in Eq. (\ref{kdec-lim1}). This allows one to derive stringent limits on the mixing matrix elements 
$U_{eh}, U_{\mu h}$. In this way an upper limit $|U_{\mu h}|^2\leq 10^{-9}$ has been derived in Ref.~\cite{Dib:2000wm}
from $\kd$ decay.

As seen from Table 1, the ratio $\Gamma_{s+t}/\Gamma_{s}$ is less than 1 below the resonant region
and greater than 1 above it. This behavior is explained by the fact that 
the interference of the s- and t-channel diagrams is destructive for $m_{\nu_h}$ below the resonant region 
and constructive above it. 
One can also notice that the ratio $\Gamma_{s+t}/\Gamma_{s}$ approaches its asymptotic value 

\begin{equation}
\label{infinity2}
\frac{\Gamma_{s+t}}{\Gamma_{s}}=
\left(\frac{4}{3}\right)^2\approx 1.78 \qquad ({m_\nu\to\infty}) 
\end{equation}
at $m_\nu\approx 10 $ GeV. This asymptotic relation follows from Eq. (\ref{infinity1}) and is independent of 
hadronic model.

\section{\label{sec:5}Summary}

We studied the hadronic structure aspects of the lepton number violating
$K^+\to \pi^- l^{+}_1 l^{+}_2$ decays within the Relativistic 
Constituent Quark Model. We considered a particular mechanism of these 
decays via Majorana neutrino exchange and derived the decay rates as functions
of neutrino mass. Our special interest was focussed on the relative contribution of 
the t-channel neutrino exchange diagram in Fig.~1(b). We have shown that it is comparable 
with the contribution of the s-channel diagram for all values of neutrino mass $m_{\nu_h}$ 
except for the resonant domains (\ref{domain}) where
the s-channel diagram Fig.~1(a) absolutely dominates in the decay rates. Outside of this domain
the relative contribution of the t-channel diagram varies between $\sim$20\% and $\sim$45\%. 
This conclusion contrasts with the previous study of Ref.~\cite{Abad:1984gh}
claiming this contribution to be always negligible in the considered decays. 
We also pointed out that the values of the decay 
rates in the resonant regions of neutrino mass represent hadronic model 
independent theoretical absolute upper bounds for the Majorana 
neutrino contribution to  $K^+\to \pi^- l^{+}_1  l^{+}_2$ decay processes.

\begin{acknowledgments}
%
This work was supported in part by Fondecyt (Chile) under
grant 1030244. M.A.I. appreciates the support from Tuebingen U.
where the part of this work has been carried out and also
the Russian Found of Basic Research under Grant No. 04-02-17370 
and the Heisenberg-Landau program.
\end{acknowledgments}

\appendix

\section{\label{sec:appA} Technical Details}

Here we present the details of the calculations leading to the expression (\ref{app-f}) for the t-channel
hadronic matrix element.

We start with the expression in Eq.~(\ref{t-ch-model}). Using the vertex functions in the form (\ref{vertex-f}) and
the $\alpha$-representation for the denominators
of quark propagators
$$
\frac{1}{m^2-K^2}=\int\limits_0^\infty d\alpha e^{-\alpha(m^2-K^2)}
$$
one can write down
\begin{eqnarray}
H^{\mu_1\mu_2}_t&=&3 g_\pi g_K
\prod\limits_{i=1}^5\int\limits_0^\infty d\alpha_i
e^{-\alpha_1 m^2_s-(\alpha_2+\alpha_3+\alpha_4)m^2_q-\alpha_5 m^2_\nu
+(\alpha_1 c_2^2+\alpha_2 c_1^2)p^2
+(\alpha_3+\alpha_4)p^{\prime 2}/4+\alpha_5 q_{12}^2}
\label{tdiag}\\
&\times&
\int\frac{d^4k_1}{(2\pi)^4i}\int\frac{d^4k_2}{(2\pi)^4i}
e^{kak+2kr}
\nonumber\\
&\times&
{\rm tr}\left\{\gamma^5 [m_s+\not\! k_1-c_2\not\! p] O^{\mu_1}
[m_u+\not\! k_2-\not\! p'/2]\gamma^5
[m_d+\not\! k_2+\not\! p'/2] O^{\mu_2}
[m_u+\not\!k_1+c_1\not\! p]\right\}
\nonumber
\end{eqnarray}
where
\[
a=\left(
\begin{array}{cc}
\mbox{$w_K+\alpha_1+\alpha_2+\alpha_5$} & \mbox{$-\alpha_5$}\\
&\\
\mbox{$-\alpha_5$} & \mbox{$w_\pi+\alpha_3+\alpha_4+\alpha_5$}
\end{array}
\right)
\]
with $w_P=1/\Lambda^2_P$ and
\[
r=\left(
\begin{array}{c}
\mbox{$(c_1\alpha_2-c_2\alpha_1)\,p+\alpha_5\, q_{12}$} \\
 \\
\mbox{$(\alpha_3-\alpha_4)\,p'/2-\alpha_5\, q_{12}$}
\end{array}
\right)
\]
Then we use the differential representation of the numerator
$$
{\rm num}(\not\! k_1, \not\! k_2)\, e^{2kr}=
{\rm num}\left(
\frac{1}{2} \not\!\partial_1,\frac{1}{2} \not\!\partial_2\right)\,
e^{2kr}
$$
with $\not\!\partial_i=\gamma^\alpha \partial/\partial r^\alpha_i$.
Integrating out the loop momenta
$$
\int\frac{d^4k_1}{(2\pi)^4i}\int\frac{d^4k_2}{(2\pi)^4i}
e^{kak+2kr}
=\frac{1}{2^8\pi^4}\frac{1}{|a|^2}e^{-ra^{-1}r}
$$
we arrive at the expression 
\begin{eqnarray}
H^{\mu_1\mu_2}(q_1,q_2;m_\nu)_t &=& \frac{3 g_\pi g_K}{2^8\pi^4}
\prod\limits_{i=1}^5\int\limits_0^\infty \frac{d\alpha_i}{|a|^2}e^{-z}
\cdot {\rm num}^{\mu_1\mu_2},
\label{tdiag1}
\end{eqnarray}
where
\begin{eqnarray}
&&{\rm num}^{\mu_1\mu_2} =
{\rm tr}\left[
\gamma^5\, [m_s-(a^{-1}\not\! r)_1-c_2\not\! p]\, O^{\mu_1}\,
[m_u-(a^{-1}\not\! r)_2 -\not\! p'/2]
\right.
\nonumber\\
&\times&
\gamma^5\,
[m_d-(a^{-1}\not\! r)_2 +\not\! p'/2]\, O^{\mu_2}\,
[m_u-(a^{-1}\not\! r)_1 +c_1\not\! p]
\nonumber\\
&-&\frac{1}{2}a^{-1}_{21}\,
\gamma^5\, \gamma^\alpha\,  O^{\mu_1}\,
\gamma^\alpha\, \gamma^5\,
[m_d-(a^{-1}\not\! r)_2 +\not\! p'/2]\, O^{\mu_2}\,
[m_u-(a^{-1}\not\! r)_1 +c_1\not\! p]
\nonumber\\
&-&\frac{1}{2}a^{-1}_{21}\,
\gamma^5\, \gamma^\alpha \, O^{\mu_1}\,
[m_u-(a^{-1}\not\! r)_2 -\not\! p'/2]\,\gamma^5\,
\gamma^\alpha\, O^{\mu_2}\,
[m_u-(a^{-1}\not\! r)_1 +c_1\not\! p]
\nonumber\\
&-&\frac{1}{2}a^{-1}_{11}\,
\gamma^5\, \gamma^\alpha\,  O^{\mu_1} \,
[m_u-(a^{-1}\not\! r)_2 -\not\! p'/2]\,\gamma^5\,
[m_d-(a^{-1}\not\! r)_2 +\not\! p'/2]\, O^{\mu_2} \,
\gamma^\alpha
\nonumber\\
&-&\frac{1}{2}a^{-1}_{22}\,
\gamma^5\, [m_s-(a^{-1}\not\! r)_1-c_2\not\! p]\, O^{\mu_1} \,
\gamma^\alpha\, \gamma^5\,
\gamma^\alpha\,  O^{\mu_2} \,
[m_u-(a^{-1}\not\! r)_1 +c_1\not\! p]
\nonumber\\
&-&\frac{1}{2}a^{-1}_{12}\,
\gamma^5\, [m_s-(a^{-1}\not\! r)_1-c_2\not\! p]\, O^{\mu_1} \,
\gamma^\alpha\, \gamma^5\,
[m_d-(a^{-1}\not\! r)_2 +\not\! p'/2]\, O^{\mu_2} \,
\gamma^\alpha
\nonumber\\
&-&\frac{1}{2}a^{-1}_{12}\,
\gamma^5\, [m_s-(a^{-1}\not\! r)_1-c_2\not\! p]\, O^{\mu_1}\,
[m_u-(a^{-1}\not\! r)_2 -\not\! p'/2]\,\gamma^5\,
\gamma^\alpha\,  O^{\mu_2}\,
\gamma^\alpha
\nonumber\\
&+&\frac{1}{4}a^{-1}_{21}a^{-1}_{12}\,
\gamma^5\,\gamma^\beta\, O^{\mu_1}\,\gamma^\beta\,\gamma^5\,
\gamma^\alpha \, O^{\mu_2}\, \gamma^\alpha
+\frac{1}{4}a^{-1}_{22}a^{-1}_{11}\,
\gamma^5\,\gamma^\beta\, O^{\mu_1}\,\gamma^\alpha\,\gamma^5\,
\gamma^\alpha\,  O^{\mu_2}\, \gamma^\beta
\nonumber\\
&+&
\left.
\frac{1}{4}a^{-1}_{12}a^{-1}_{21}\,
\gamma^5\,\gamma^\beta\, O^{\mu_1}\,\gamma^\alpha\,\gamma^5\,
\gamma^\beta \, O^{\mu_2} \,\gamma^\alpha
\right]
\nonumber
\end{eqnarray}
and 
\begin{eqnarray}
z &=&\alpha_1\, m^2_s+(\alpha_2+\alpha_3+\alpha_4)\,m^2_q+\alpha_5\, m^2_\nu
\nonumber\\
&-&(\alpha_1 c_2^2+\alpha_2 c_1^2)\,p^2
 -(\alpha_3+\alpha_4)\,p^{\prime 2}/4-\alpha_5\, q_{12}^2
+ra^{-1}r.
\end{eqnarray}
Thus, we have reduced the two-loop integrations to the 5-fold integrals
over $\alpha$-parameters. We use the FORM code \cite{Vermaseren:2000nd} for the calculation of the trace and 
end up with the ten independent Lorentz structures
\begin{eqnarray}\nonumber
H^{\mu_1\mu_2}(q_1,q_2;m_\nu)_t &=&
 H_t^{g}(s_1,s_2;m_{\nu})\,g^{\mu_1\mu_2}
+H_t^{pp}(s_1,s_2;m_{\nu})\,p^{\mu_1}\,p^{\mu_2}
+H_t^{p'p'}(s_1,s_2;m_{\nu})\,p^{\prime\,\mu_1}\,p^{\prime\,\mu_2}
\label{tensor}\\
&+&
 H_t^{qq}(s_1,s_2;m_{\nu})\,q_{12}^{\mu_1}\,q_{12}^{\mu_2}
+H_t^{p'p}(s_1,s_2;m_{\nu})\,
\left(p^{\mu_1}\,p^{\prime\,\mu_2}+p^{\prime\,\mu_1}\,p^{\mu_2}\right)
\nonumber\\
&+&
 H_t^{pq}(s_1,s_2;m_{\nu})\,
\left(p^{\mu_1}\,q_{12}^{\mu_2}+q_{12}^{\mu_1}\,p^{\mu_2}\right)
+H_t^{p'q}(s_1,s_2;m_{\nu})\,
\left(p^{\prime\,\mu_1}\,q_{12}^{\mu_2}+q_{12}^{\mu_1}\,p^{\prime\,\mu_2}\right)
\nonumber\\
&+&
H_t^{epp'}(s_1,s_2;m_{\nu})\,
i\varepsilon^{\rho_1\rho_2\mu_1\mu_2}\,p_{\rho_1}\,p'_{\rho_2}
+H_t^{epq}(s_1,s_2;m_{\nu})\,
i\varepsilon^{\rho_1\rho_2\mu_1\mu_2}\,p_{\rho_1}\,q_{12\,\rho_2}
\nonumber\\
&+&
H_t^{ep'q}(s_1,s_2;m_{\nu})\,
i\varepsilon^{\rho_1\rho_2\mu_1\mu_2}\,p'_{\rho_1}\,q_{12\,\rho_2}.
\nonumber
\end{eqnarray}
We have shown that 
\begin{eqnarray} \label{relations}
H_t^{pp}   = H_t^{qq}= H_t^{pq}= H_t^{epq}\equiv 0,\ \ 
H_t^{epp'} = H_t^{p'p}, \ \ 
H_t^{ep'q}  = -H_t^{p'q}.
\nonumber
\end{eqnarray}
The function $H_t^{p'p'}$ has been numerically found to be
negligibly small.
We calculated the three remaining structure functions $H^A_t(s_1,s_2)$
($A=g,p'p,p'q$) using the Fortran code and then approximated
them by the functions
\begin{equation}
\label{approx}
H^A_t(s_1,s_2;m_{\nu})=\frac{H(s_1^{\rm min},s_2^{\rm min})}
           {1+b_1 x_1+b_2 x_2+b_{11}x_1^2+b_{22}x_2^2++b_{12}x_1x_2},
\end{equation}
with  $x_i=s_i-s_i^{\rm min}$ (i=1,2). For $\kd$ one has 
$s_1^{\rm min}=4\,m^2_\mu$, $s_2^{\rm min}=(m_\pi+m_\mu)^2$.
The coefficients $b_1,b_2 ,b_{11},b_{22},b_{12}$ and
$H(s_1^{\rm min},s_2^{\rm min})$ depend on neutrino mass $m_{\nu}$. 
The code for their numerical calculations is available from the authors.

\begin{table}[h]
\begin{center}
\def\arraystretch{2.0}
\caption{The total $\kd$ decay rate $\Gamma_{s+t}$ and the ratio 
$\Gamma_{s+t}/\Gamma_s$ vs. neutrino mass $m_{\nu_h}$.}
\begin{tabular}{|c|c|c||c|c|c||c|c|c||c|c|c|}

\hline
 $m_{\nu_h}$ &$\Gamma_{s+t}/|U_{\mu h}|^4$ & $\,\,\frac{\Gamma_{s+t}}{\Gamma_s}$\,\, &
 $m_{\nu_h}$ &$\Gamma_{s+t}/|U_{\mu h}|^4$ & $\,\,\frac{\Gamma_{s+t}}{\Gamma_s}$\,\, &
 $m_{\nu_h}$ &$\Gamma_{s+t}/|U_{\mu h}|^4$ & $\,\,\frac{\Gamma_{s+t}}{\Gamma_s}$\,\, &
 $m_{\nu_h}$ &$\Gamma_{s+t}/|U_{\mu h}|^4$ & $\,\,\frac{\Gamma_{s+t}}{\Gamma_s}$\,\, \\[-3mm]
(eV)&(GeV)&& (KeV)&(GeV)&& (MeV)&(GeV)&& (GeV)&(GeV)&\\
\hline
 1   &$ 0.65 \cdot 10^{-48} $& 0.85  & 1   &$ 0.65 \cdot 10^{-42} $&0.85 & 1   &$ 0.65 \cdot 10^{-36} $& 0.85 & 1  &$ 0.13 \cdot 10^{-31}$& 1.33
 \\
 250 &$ 0.4 \cdot 10^{-43} $& 0.85  & 250 &$ 0.40 \cdot 10^{-37} $&0.85 & 250 &$ 0.6 \cdot 10^{-18} $& 1.00 & 250&$ 0.22 \cdot 10^{-36}$& 1.78
 \\
 500 &$ 0.16 \cdot 10^{-42} $& 0.85  & 500 &$ 0.16 \cdot 10^{-36} $&0.85 & 500 &$ 0.10 \cdot 10^{-30} $& 1.00 & 500&$ 0.55 \cdot 10^{-37}$& 1.78
 \\
 750 &$ 0.36 \cdot 10^{-42} $& 0.85  & 750 &$ 0.36 \cdot 10^{-36} $&0.85 & 750 &$ 0.2 \cdot 10^{-31} $& 1.25 & 750&$ 0.25 \cdot 10^{-37}$& 1.78
 \\
\hline
\end{tabular}
\end{center}
\end{table}


\newpage

\begin{thebibliography}{999}
%
\bibitem{seesaw} P.~Minkowski, Phys. Lett. B {\bf 67},  421 (1977); T.~Yanagida, proceedings of 
{\it the Workshop on Unified Theories and Baryon 
Number in the Universe}, Tsukuba, 1979, eds. A.~Sawada, A.~Sugamoto, KEK Report No. 79-18, Tsukuba; 
S.~Glashow, in {\it Quarks and Leptons, Carg\'ese} 1979, eds. M.~L\'evy, {\it et al.}, (Plenum, 1980, New York); 
M.~Gell-Mann, \mbox{P.~Ramond}, \mbox{R.~Slansky}, proceedings of {\it the Supergravity Stony Brook Workshop}, New York, 1979, 
eds. \mbox{P.~Van Niewenhuizen}, D.~Freeman (North-Holland, Amsterdam); 
R.~Mohapatra, \mbox{G.~Senjanovi\'c}, Phys.Rev.Lett. {\bf 44},  912 (1980). 
%
\bibitem{seesaw-recent}
R.~N.~Mohapatra, arXiv:hep-ph/0211252; G. Altarelli and F. Feruglio, arXiv:hep-ph/0405048;
C.~H.~Albright, Int. J. Mod. Phys. A {\bf 18}, 3947 (2003); J. C. Pati, arXiv:hep-ph/0407220;
G.~Senjanovi\'c, arXiv:hep-ph/0501244.
%
\bibitem{rod02} W.~Rodejohann, J. Phys. G {\bf 28}, 1477 (2002).
%
\bibitem{Dib:2000ce}
C.~Dib, V.~Gribanov, S.~Kovalenko and I.~Schmidt,
Part.\ Nucl.\ Lett.\  {\bf 106}, 42 (2001).
%
\bibitem{kla01a} H.~V.~Klapdor-Kleingrothaus, {\it et~al.}, Eur. Phys. J. A {\bf 12}, 147 (2001).
%
\bibitem{aal02} C.~E.~Aalseth et~al. (collaboration 16EX), Phys. Rev. D {\bf 65}, 092007 (2002).
%
\bibitem{arn04} C.~Arnaboldi et~al., Phys. Lett. B {\bf 584}, 260 (2004).
%
\bibitem{PDG} Review of Particle Physics, Phys. Lett. B {\bf 592}, 1 (2004).  
%
\bibitem{dorokhov:2002} A.~Belyaev, M.~Chizhov, A.~Dorokhov, J.~R.~Ellis, M.~E.~Gomez and S.~Lola,
Eur.\ Phys.\ J.\ C {\bf 22}, 715 (2002). 
\bibitem{Dib:2000wm}
C.~Dib, V.~Gribanov, S.~Kovalenko and I.~Schmidt,
Phys.\ Lett.\ B {\bf 493}, 82 (2000).
%
\bibitem{Ng:1978ij}
J.~N.~Ng and A.~N.~Kamal,
Phys.\ Rev.\ D {\bf 18}, 3412 (1978).
%
\bibitem{Abad:1984gh}
J.~Abad, J.~G.~Esteve and A.~F.~Pacheco,
Phys.\ Rev.\ D {\bf 30}, 1488 (1984).
%
\bibitem{Kdec3}
L.~S.~Littenberg and R.~E.~Shrock,
Phys.\ Rev.\ Lett.\  {\bf 68}, 443 (1992);
Phys.\ Lett.\ B {\bf 491}, 285 (2000).
%
%
\bibitem{model}
M.~A.~Ivanov, M.~P.~Locher and V.~E.~Lyubovitskij,
Few Body Syst.  {\bf 21}, 131 (1996);
%
M.~A.~Ivanov, J.~G.~K\"{o}rner and P.~Santorelli,
Phys. Rev. D {\bf 63}, 074010 (2001);
%
A.~Faessler, T.~Gutsche, M.~A.~Ivanov, J.~G.~K\"{o}rner and V.~E.~Lyubovitskij,
Eur.\ Phys.\ J.\ C {\bf 4}, 18 (2002).
%
\bibitem{GKS:2001}
V.~Gribanov, S.~Kovalenko and I.~Schmidt,
Nucl.Phys. B {\bf 607}, 355 (2001). 
%
%
\bibitem{Vermaseren:2000nd}
J.~A.~M.~Vermaseren,
arXiv:math-ph/0010025.
\end{thebibliography}
\end{document}